\documentclass[10pt]{iopart}
\begin{document}

\title[Strangeness Content in the Nucleon]{Strangeness Content in the
Nucleon}
 
\author{Keh-Fei Liu \footnote{liu@pa.uky.edu \\
Invited talk presented at V International Conference on Strangeness in
Quark Matter, ``Strangeness 2000'', Berkeley, CA, July 20--25, 2000}}

\address{Department of Physics, University of Kentucky, Lexington, KY
40506, USA}

\begin{abstract}
I will review recent studies of strangeness content
in the nucleon pertaining to the flavor-singlet $g_A^0$, the $\bar{s}s$
matrix element and the strangeness electric and magnetic form factors 
$G_E^s(q^2)$ and $G_M^s(q^2)$, based on lattice QCD
calculations. I shall also discuss the relevance of incorporating the strangeness
content in nuclei in regard to strange baryon-antibaryon productions from
proton-nucleus and nucleus-nucleus collisions at SPS and RHIC energies.
\end{abstract}




\section{Introduction}
There has been a great deal of interest of late in measuring and understanding
the strangeness content of the nucleon. This includes the strangeness magnetic
moment~\cite{bec00,SAMPLE00} and electric form factor~\cite{ceb91}, the
$\bar{s}s$ matrix element and $\pi N \sigma$ term~\cite{hol99}, the strange
quark polarization~\cite{hv99} and orbital angular momentum~\cite{mdl00}, and
the strange to non-strange sea ratio in deep inelastic
scattering~\cite{NuTev99}. Why do we concern ourselves with
strangeness in the nucleon at all? Why is there a surge of interest in this
subject recently? Is it relevant to the
production of strange and anti-strange hadrons in proton-nucleus and
nucleus-nucleons collisions at relativistic energies which, of course, is the
focus of this conference? These are the issues that I want to discuss in this
talk.

The conventional wisdom based on the valence quark model
picture suggests that there is no need to consider strangeness in the
nucleon. This is so because valence quark models have been
successful in describing hadron spectroscopy~\cite{dgg75,lw78b,ik78,cjj74} and 
baryon magnetic moments~\cite{dgg75,br88,pon96}, especially the neutron to
proton magnetic moment ratio $\mu_n/\mu_p$. It is also successful in
delineating the pattern of electro-magnetic~\cite{clo79,ono76,ki80},
semileptonic and nonleptonic weak decays~\cite{dgh92}. In particular,
up until late eighties, there had not been a prevailing reason to question
the validity of the Okubo-Zweig-Iizuka (OZI) rule~\cite{ozi63} which seemed 
to work well in the available experiments. Thus, there is no compelling reason
to include the higher Fock space such as $s\bar{s}$ pair in the nucleon 
wavefunction beyond the valence $u$ and $d$ quarks.

On the other hand, there have been hints all along that the
valence quark model for the light hadrons is incomplete and inadequate 
in elucidating a number of quantities and relations. Examples abound, they
include the $\eta'$ meson mass -- the $U(1)$ anomaly, the vector dominance, 
the Goldberger-Treiman relation and the $\pi N \sigma$ term puzzle.
These examples can be understood in terms of sea quarks
and meson clouds via models and current algebra based on chiral symmetry, an
inherent symmetry in QCD. However, these warnings did not amount to a
wake up call about the shortcomings of the valence quark model until  
the discovery of the
large and negative polarization of the strange quark in polarized deep
inelastic scattering~\cite{emc88,e14395,smc97} which led to a small flavor-singlet
$g_A^0$, about $\frac{1}{4}/\frac{1}{3}$ of that
predicted by the non-relativistic/relativistic quark model. This discrepancy
has not been satisfactorily explained in any hadronic model and has been
dubbed the `proton spin crisis'.  It is mostly
understood only through direct calculations in lattice QCD~\cite{dll95,
fko95b,guv99}. This, I believe, is a watershed in hadron physics. From this
point on,  people are forced to take the intrinsic strangeness, and sea quarks
for that matter, in the nucleon seriously. 

I shall illustrate
the role of strangeness in three examples, namely the flavor-singlet
$g_A^0$, the $\langle N|\bar{s}s|N\rangle$ matrix element, and the strangeness
magnetic moment and electric form factor. I shall mainly discuss what we
have learned from the available lattice QCD calculations. 
Lattice QCD offers an {\it ab initio\/} method of computing the
non-perturbative properties of the strong interaction; it can account for all
characteristics of nucleon and hadron properties to arbitrary accuracy, in
principle.  In practice, it is subject to great technical challenges which
however are being addressed with increasingly sophisticated computational,
algorithmic and theoretical ideas~\cite{lat99}.  An important goal has been to
make a quantitative test of QCD --- to demonstrate that the underlying dynamics
of quarks and gluons, as specified by QCD, can describe accurately the
properties, as observed in the laboratory, of nucleons and other hadrons.  This
has not been achieved yet, but as numerical simulations of increasing size and
sophistication are applied to this problem, we see QCD being able to account
with increasing detail for the structure of the hadronic world.
 
For the short term, it proves to be beneficial to carry out quenched lattice
calculations to gain qualitative and semi-quantitative understanding on the physical
quantities that are not reliably obtainable from the present models.  Model
predictions on these quantities can differ by as much as a factor of $3$ to $5$
(e.g.\ the monopole mass in the $\pi NN$ form factor, the strangeness content
$\bar{s}s$ in the nucleon, and the $E2/M1$ ratio in $N\gamma \leftrightarrow
\Delta$) and sometimes differ even in sign (e.g.\ the strange magnetic moment
and electric form factor in the nucleon).  Even though our present lattice
calculation based on the Wilson fermion in the quenched approximation may have
a systematic error as large as 20\% for the connected insertion (valence and
connected sea) part and maybe larger for the disconnected insertion
(disconnected sea) part, it is already better than the systematic errors spanned
by the model dependence. Furthermore, the systematic errors of these quantities
can be put under control when the study of the large volume, continuum limit,
and chiral limit are included.  These systematic errors will be addressed with
Neuberger's overlap fermion~\cite{neu98} which has lattice chiral symmetry and
good scaling behavior. In addition, it admits numerical calculation close to
the physical quark masses (e.g. physical pion mass at 200 MeV) without
significant critical slowing down~\cite{dll00}. Systematic errors due to the
dynamical fermion effect will have to be addressed when enough computer
resources are available. The present day dynamical fermion simulations are
still performed with quite heavy quarks which may not reflect the realistic
quark loop effects beyond a shift in coupling constant like a dielectric
constant effect~\cite{lw99}.

\section{Strangeness Content in the Nucleon}

\subsection{Flavor-singlet $g_A^0$ -- Quark Spin Content of the Nucleon}

The polarized DIS experiments~\cite{emc88,e14395,smc97} 
found a surprisingly small flavor-singlet axial coupling constant 
$g_A^0$ (0.27(10)~\cite{e14395} and 0.28(16)~\cite{smc97}). Being the quark 
spin content of the nucleon; i.e. $g_A^0 = \Delta u + \Delta d + \Delta s$,
this is much smaller than the expected value of 
unity from the non-relativistic quark model or 0.75 from the $SU(6)$ relation 
(i.e. 3/5 of the isovector coupling $g_A^3 = 1.2574$). This has attracted
a lot of theoretical attention~\cite{cheng96} and the ensuing confusion 
was dubbed the ``proton spin crisis''.

Direct lattice calculations of $g_A^0$ from the forward matrix element of
the flavor-singlet axial current has been carried out and 
the smallness of $g_A^0$ is understood~\cite{dll95,fko95b,guv99}.  
As a flavor-singlet quantity, $g_A^0$ is composed of two components.
One is from the connected insertion (C.I.) and the other is from the
disconnected insertion (D.I.). Therefore,
$g_A^0 = g_A^0 (C. I.) + g_A^0 (D. I.)$
where $g_A^0 (C. I.)$ is obtained from the connected insertion in Fig.
1(a) and $g_A^0 (D. I.)$ is obtained from the disconnected insertion 
in Fig. 1(b). Lattice calculation~\cite{dll95} indicates that
each of the $u, d,$ and $s$ flavors contributes $ - 0.12 \pm 0.01$ to the
D.I. (Fig. 1(b)). This negative vacuum polarization from the disconnected
sea quarks is largely responsible for bringing the value of $g_A^0$ from 
$ g_A^0 (C. I.) = 0.62 \pm 0.09$ to $0.25 \pm 0.12$, in agreement with the
experimental value (See Table 1). This is an example where the disconnected sea
contributes substantially and leads to a large breaking in the $SU(6)$
relation. Thus, it is understandable that it should come as a
surprise to the valence quark model --- the latter does not have the sea degree
of freedom, disconnected or connected, and has simply ignored it by assuming the
OZI rule.

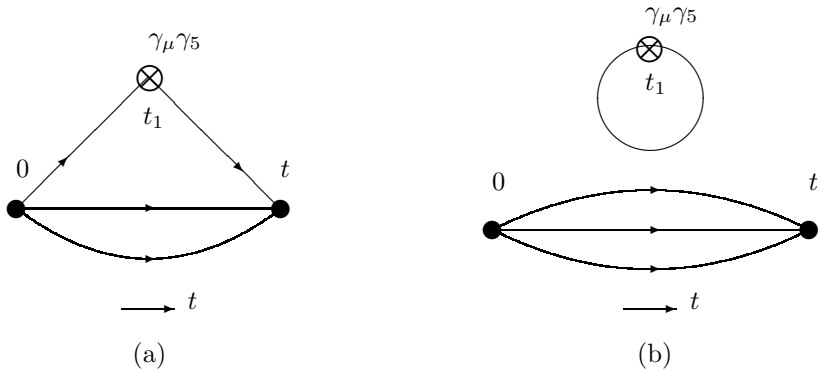
\begin{figure}[ht]
\hspace*{2.0in}
\setlength{\unitlength}{0.01pt}
\begin{picture}(45000,20000)
\put(-10000, 8800){\circle*{700}}
\put(-10000, 8800){\line(1,1){5000}}
\put(-8232, 10568){\vector(1,1){200}}
\put(0000, 8800){\line(-1,1){5000}}
\put(-1591,10391){\vector(1,-1){200}}
\put(-5500, 13500){{\bf $\bigotimes$}}
\put(-10000, 8800){\line(1,0){10000}}
\put(-5000,  8800){\vector(1,0){300}}
\put(0000, 8800){\circle*{700}}
\qbezier(-10000, 8800)(-5000, 5000)(00, 8800)
\put(-5000, 6900){\vector(1,0){300}}
\put(-6000, 5000){\vector(1,0){2000}}
\put(-5000,15000){{\bf $\gamma_\mu \gamma_5$}}
\put(-5200,12000){$t_1$}
\put(-3500, 5000){$t$}
\put(-10000,10000){$0$}
\put(0,10000){$t$}
\put(-5600, 3000){(a)}
\put(8000, 8000){\circle*{700}}
\put(8000, 8000){\line(1,0){12000}}
\put(14000, 8000){\vector(1,0){400}}
\put(20000, 8000){\circle*{700}}
\qbezier(8000, 8000)(14000, 5000)(20000, 8000)
\put(14000, 6500){\vector(1,0){400}}
\qbezier(8000, 8000)(14000,11000)(20000, 8000)
\put(14000, 9500){\vector(1,0){400}}
\put(14000,13000){\circle{4000}}
\put(13400,14600){{\bf $\bigotimes$}}
\put(13800,16000){{\bf $\gamma_\mu \gamma_5$}}
\put(13000, 5000){\vector(1,0){2000}}
\put(13800,13300){$t_1$}
\put(15500, 5000){$t$}
\put(8000, 9500){$0$}
\put(20000, 9500){$t$}
\put(13500, 3000){(b)}
%
\end{picture}
\caption{Quark line diagrams of the three-point function in the Euclidean path 
integral formalism for evaluating $g_A^0$ from the flavor-singlet
axial-vector current. (a) is the connected insertion which contains the
valence and connected sea degrees of freedom and (b) is the disconnected
insertion which contains the disconnected quark sea.} 
\end{figure}

\begin{table}[ht]
\caption{Axial coupling constants and quark spin contents of proton in
comparison with experiments, the non-relativistic quark model (NRQM), and
the relativistic quark model (RQM).}
\begin{tabular}{llllll}
 \multicolumn{1}{c}{} &\multicolumn{1}{c}{C. I.} &\multicolumn{1}{c}{C. I.
 + D. I.} &  \multicolumn{1}{c} {Experiments} &  \multicolumn{1}{c} 
 {NRQM} &\multicolumn{1}{c} {RQM}\\
 \hline
 $g_A^0 {\scriptstyle =\Delta u + \Delta d + \Delta s}$ 
 &  0.62(9) & 0.25(12)& 0.28(16) \cite{smc97}, 0.27(10)\cite{e14395}
 & 1 & 0.754  \\
 $g_A^3 {\scriptstyle =\Delta u - \Delta d}$ & 1.20(10) \cite{ldd94a} 
 &1.20(10)& 1.2573(28) & 5/3 & 1.257 \\
 $g_A^8 {\scriptstyle =\Delta u + \Delta d - 2\Delta s}$ & 0.62(9) & 
 0.61(13) &  0.579(25) \cite{cr93} & 1 & 0.754 \\
 $\Delta u $ & 0.91(10) & 0.79(11) & 0.82(5)\cite{smc97}, 0.82(6)\cite{e14395}
  & 4/3 & 1.01 \\
 $\Delta d $ & -.29(10) &- 0.42(11) &  - 0.44(5)\cite{smc97}, - 0.44(6)
 \cite{e14395} & -1/3 & -0.251  \\
 $\Delta s $    &    & - 0.12(1) & - 0.10(5)\cite{smc97}, - 0.10(4)
 \cite{e14395} & 0 & 0 \\
 $F_A {\scriptstyle = (\Delta u - \Delta s)/2}$ & 0.45(6)& 0.45(6) & 0.459(8) 
 \cite{cr93} & 2/3 & 0.503 \\
 $D_A {\scriptstyle = (\Delta u - 2 \Delta d + \Delta s)/2}$ & 0.75(11)
 & 0.75(11) & 0.798(8) \cite{cr93} & 1 & 0.754\\
 $F_A/D_A$ & 0.60(2) & 0.60(2) & 0.575(16) \cite{cr93} & 2/3 & 2/3  \\
 \hline
\end{tabular}
\end{table}

We see in Table 1 that the calculated result of the strange polarization
$\Delta s$ is in good agreement with the experiments. Since the contribution 
from the strange quark comes only from the disconnected sea in Fig. 1(b)
~\cite{liu00}, the role of the strange and the disconnected sea contribution 
from $u$ and $d$ quarks is clear and well-defined in the three-point function 
as illustrated in Fig. 1(b). From the path-integral formulation of the hadronic
tensor,  it is learned~\cite{liu00, ldd99} that there is connected sea in the
form of the non-perturbative Z-graphs in the C.I. represented by Fig. 1(a)
which is responsible for the violation of the Gottfried sum rule, for example. 
Although the strange quark does not contribute to the connected sea in
Fig. 1(b), it is still interesting to find out what effect the connected
sea has on the physical quantities we are studying. 
Since the connected sea contribution to the C.I. of three-point functions is
entangled with the valence, we can not separate it out as is done for the 
disconnected sea. To see its effect indirectly, we consider the ratio
\begin{eqnarray}   \label{R_A}
R_A &=&\frac{g_A^0}{g_A^3} = \frac{\Delta u + \Delta d + \Delta s}
{\Delta u - \Delta d}   \nonumber \\
&= &\frac{(\Delta u + \Delta d)(C.I.) + (\Delta u + \Delta d + \Delta s)(D.I.)}
{\Delta u - \Delta d}
\end{eqnarray}
as a function of the quark mass. Our results which correspond
to the range between strange and twice the charm masses are plotted in Fig. 2
as a function of the quark mass $ma = ln (4\kappa_c/\kappa -3)$.
The dotted line is the valence quark model prediction of 3/5 for both
the non-relativistic and relativistic cases. For heavy quarks
(i.e. $\kappa \geq 0.133$ or $ma \geq 0.4$ in Fig. 2), we see that 
the ratio $R_A$ is 3/5 irrespective of whether the D.I. is included	
(shown as $\bullet$ in Fig. 2) or not (C.I. alone is indicated as
$\circ$). This is to
be expected because the disconnected/connected sea quarks which are pair produced
via the Z-graphs/loops are suppressed for non-relativistic quarks 
by $O(p/m_q)$ or $ O(v/c)$. As for light quarks, the full result
(C.I. + D.I.) is much smaller than 3/5 largely due to the negatively
polarized sea contribution in the D.I. (NB. Table 1 lists the results at
the chiral limit.)  Even for the C.I. alone, 
$R_A$ still dips under 3/5. As we shall see later this is caused
by the disconnected sea quarks. 

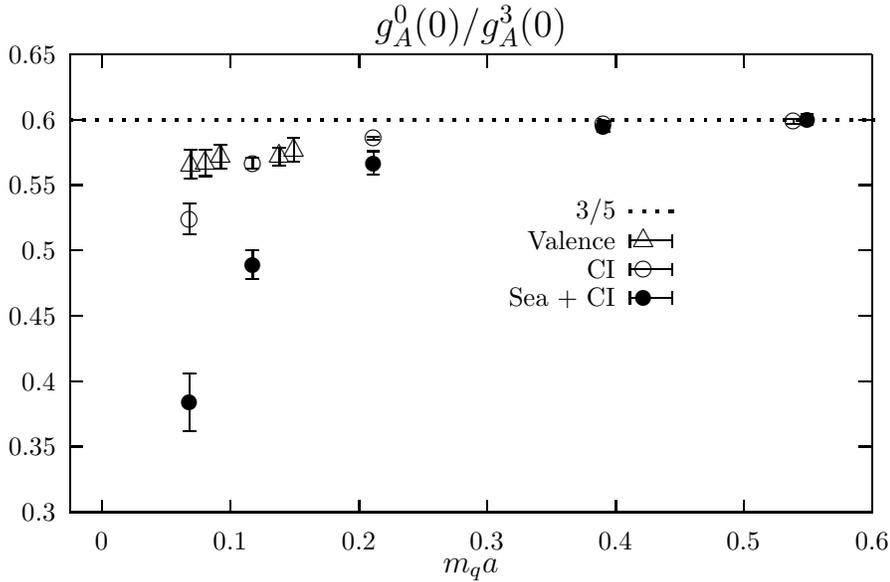
\begin{figure}[h]
\setlength{\unitlength}{0.240900pt}
\ifx\plotpoint\undefined\newsavebox{\plotpoint}\fi
\sbox{\plotpoint}{\rule[-0.200pt]{0.400pt}{0.400pt}}%
\begin{picture}(1500,900)(0,0)
\font\gnuplot=cmr10 at 10pt
\gnuplot
\sbox{\plotpoint}{\rule[-0.200pt]{0.400pt}{0.400pt}}%
\put(176.0,113.0){\rule[-0.200pt]{4.818pt}{0.400pt}}
\put(154,113){\makebox(0,0)[r]{0.3}}
\put(1416.0,113.0){\rule[-0.200pt]{4.818pt}{0.400pt}}
\put(176.0,216.0){\rule[-0.200pt]{4.818pt}{0.400pt}}
\put(154,216){\makebox(0,0)[r]{0.35}}
\put(1416.0,216.0){\rule[-0.200pt]{4.818pt}{0.400pt}}
\put(176.0,318.0){\rule[-0.200pt]{4.818pt}{0.400pt}}
\put(154,318){\makebox(0,0)[r]{0.4}}
\put(1416.0,318.0){\rule[-0.200pt]{4.818pt}{0.400pt}}
\put(176.0,421.0){\rule[-0.200pt]{4.818pt}{0.400pt}}
\put(154,421){\makebox(0,0)[r]{0.45}}
\put(1416.0,421.0){\rule[-0.200pt]{4.818pt}{0.400pt}}
\put(176.0,524.0){\rule[-0.200pt]{4.818pt}{0.400pt}}
\put(154,524){\makebox(0,0)[r]{0.5}}
\put(1416.0,524.0){\rule[-0.200pt]{4.818pt}{0.400pt}}
\put(176.0,627.0){\rule[-0.200pt]{4.818pt}{0.400pt}}
\put(154,627){\makebox(0,0)[r]{0.55}}
\put(1416.0,627.0){\rule[-0.200pt]{4.818pt}{0.400pt}}
\put(176.0,729.0){\rule[-0.200pt]{4.818pt}{0.400pt}}
\put(154,729){\makebox(0,0)[r]{0.6}}
\put(1416.0,729.0){\rule[-0.200pt]{4.818pt}{0.400pt}}
\put(176.0,832.0){\rule[-0.200pt]{4.818pt}{0.400pt}}
\put(154,832){\makebox(0,0)[r]{0.65}}
\put(1416.0,832.0){\rule[-0.200pt]{4.818pt}{0.400pt}}
\put(226.0,113.0){\rule[-0.200pt]{0.400pt}{4.818pt}}
\put(226,68){\makebox(0,0){0}}
\put(226.0,812.0){\rule[-0.200pt]{0.400pt}{4.818pt}}
\put(428.0,113.0){\rule[-0.200pt]{0.400pt}{4.818pt}}
\put(428,68){\makebox(0,0){0.1}}
\put(428.0,812.0){\rule[-0.200pt]{0.400pt}{4.818pt}}
\put(630.0,113.0){\rule[-0.200pt]{0.400pt}{4.818pt}}
\put(630,68){\makebox(0,0){0.2}}
\put(630.0,812.0){\rule[-0.200pt]{0.400pt}{4.818pt}}
\put(831.0,113.0){\rule[-0.200pt]{0.400pt}{4.818pt}}
\put(831,68){\makebox(0,0){0.3}}
\put(831.0,812.0){\rule[-0.200pt]{0.400pt}{4.818pt}}
\put(1033.0,113.0){\rule[-0.200pt]{0.400pt}{4.818pt}}
\put(1033,68){\makebox(0,0){0.4}}
\put(1033.0,812.0){\rule[-0.200pt]{0.400pt}{4.818pt}}
\put(1234.0,113.0){\rule[-0.200pt]{0.400pt}{4.818pt}}
\put(1234,68){\makebox(0,0){0.5}}
\put(1234.0,812.0){\rule[-0.200pt]{0.400pt}{4.818pt}}
\put(1436.0,113.0){\rule[-0.200pt]{0.400pt}{4.818pt}}
\put(1436,68){\makebox(0,0){0.6}}
\put(1436.0,812.0){\rule[-0.200pt]{0.400pt}{4.818pt}}
\put(176.0,113.0){\rule[-0.200pt]{303.534pt}{0.400pt}}
\put(1436.0,113.0){\rule[-0.200pt]{0.400pt}{173.207pt}}
\put(176.0,832.0){\rule[-0.200pt]{303.534pt}{0.400pt}}
\put(806,23){\makebox(0,0){{\large\bf $m_q a$}}}
\put(806,877){\makebox(0,0){{\Large\bf $g_A^0(0)/g_A^3(0)$}}}
\put(176.0,113.0){\rule[-0.200pt]{0.400pt}{173.207pt}}
\sbox{\plotpoint}{\rule[-0.500pt]{1.000pt}{1.000pt}}%
\put(1033,585){\makebox(0,0)[r]{3/5}}
\multiput(1055,585)(20.756,0.000){4}{\usebox{\plotpoint}}
\put(1121,585){\usebox{\plotpoint}}
\put(176,729){\usebox{\plotpoint}}
\multiput(176,729)(20.756,0.000){61}{\usebox{\plotpoint}}
\put(1436,729){\usebox{\plotpoint}}
\sbox{\plotpoint}{\rule[-0.200pt]{0.400pt}{0.400pt}}%
\put(1033,540){\makebox(0,0)[r]{Valence}}
\put(1077,540){\makebox(0,0){$\triangle$}}
\put(366,659){\makebox(0,0){$\triangle$}}
\put(389,661){\makebox(0,0){$\triangle$}}
\put(413,672){\makebox(0,0){$\triangle$}}
\put(505,672){\makebox(0,0){$\triangle$}}
\put(528,682){\makebox(0,0){$\triangle$}}
\put(1055.0,540.0){\rule[-0.200pt]{15.899pt}{0.400pt}}
\put(1055.0,530.0){\rule[-0.200pt]{0.400pt}{4.818pt}}
\put(1121.0,530.0){\rule[-0.200pt]{0.400pt}{4.818pt}}
\put(366.0,637.0){\rule[-0.200pt]{0.400pt}{10.840pt}}
\put(356.0,637.0){\rule[-0.200pt]{4.818pt}{0.400pt}}
\put(356.0,682.0){\rule[-0.200pt]{4.818pt}{0.400pt}}
\put(389.0,641.0){\rule[-0.200pt]{0.400pt}{9.877pt}}
\put(379.0,641.0){\rule[-0.200pt]{4.818pt}{0.400pt}}
\put(379.0,682.0){\rule[-0.200pt]{4.818pt}{0.400pt}}
\put(413.0,653.0){\rule[-0.200pt]{0.400pt}{8.913pt}}
\put(403.0,653.0){\rule[-0.200pt]{4.818pt}{0.400pt}}
\put(403.0,690.0){\rule[-0.200pt]{4.818pt}{0.400pt}}
\put(505.0,657.0){\rule[-0.200pt]{0.400pt}{6.986pt}}
\put(495.0,657.0){\rule[-0.200pt]{4.818pt}{0.400pt}}
\put(495.0,686.0){\rule[-0.200pt]{4.818pt}{0.400pt}}
\put(528.0,664.0){\rule[-0.200pt]{0.400pt}{8.913pt}}
\put(518.0,664.0){\rule[-0.200pt]{4.818pt}{0.400pt}}
\put(518.0,701.0){\rule[-0.200pt]{4.818pt}{0.400pt}}
\put(1033,495){\makebox(0,0)[r]{CI}}
\put(1077,495){\circle{24}}
\put(364,573){\circle{24}}
\put(463,661){\circle{24}}
\put(653,701){\circle{24}}
\put(1014,723){\circle{24}}
\put(1312,727){\circle{24}}
\put(1055.0,495.0){\rule[-0.200pt]{15.899pt}{0.400pt}}
\put(1055.0,485.0){\rule[-0.200pt]{0.400pt}{4.818pt}}
\put(1121.0,485.0){\rule[-0.200pt]{0.400pt}{4.818pt}}
\put(364.0,549.0){\rule[-0.200pt]{0.400pt}{11.804pt}}
\put(354.0,549.0){\rule[-0.200pt]{4.818pt}{0.400pt}}
\put(354.0,598.0){\rule[-0.200pt]{4.818pt}{0.400pt}}
\put(463.0,653.0){\rule[-0.200pt]{0.400pt}{4.095pt}}
\put(453.0,653.0){\rule[-0.200pt]{4.818pt}{0.400pt}}
\put(453.0,670.0){\rule[-0.200pt]{4.818pt}{0.400pt}}
\put(653.0,698.0){\rule[-0.200pt]{0.400pt}{1.204pt}}
\put(643.0,698.0){\rule[-0.200pt]{4.818pt}{0.400pt}}
\put(643.0,703.0){\rule[-0.200pt]{4.818pt}{0.400pt}}
\put(1014.0,719.0){\rule[-0.200pt]{0.400pt}{1.927pt}}
\put(1004.0,719.0){\rule[-0.200pt]{4.818pt}{0.400pt}}
\put(1004.0,727.0){\rule[-0.200pt]{4.818pt}{0.400pt}}
\put(1312.0,723.0){\rule[-0.200pt]{0.400pt}{1.927pt}}
\put(1302.0,723.0){\rule[-0.200pt]{4.818pt}{0.400pt}}
\put(1302.0,731.0){\rule[-0.200pt]{4.818pt}{0.400pt}}
\put(1033,450){\makebox(0,0)[r]{Sea + CI}}
\put(1077,450){\circle*{24}}
\put(364,286){\circle*{24}}
\put(463,501){\circle*{24}}
\put(653,661){\circle*{24}}
\put(1014,719){\circle*{24}}
\put(1334,729){\circle*{24}}
\put(1055.0,450.0){\rule[-0.200pt]{15.899pt}{0.400pt}}
\put(1055.0,440.0){\rule[-0.200pt]{0.400pt}{4.818pt}}
\put(1121.0,440.0){\rule[-0.200pt]{0.400pt}{4.818pt}}
\put(364.0,240.0){\rule[-0.200pt]{0.400pt}{21.922pt}}
\put(354.0,240.0){\rule[-0.200pt]{4.818pt}{0.400pt}}
\put(354.0,331.0){\rule[-0.200pt]{4.818pt}{0.400pt}}
\put(463.0,479.0){\rule[-0.200pt]{0.400pt}{10.840pt}}
\put(453.0,479.0){\rule[-0.200pt]{4.818pt}{0.400pt}}
\put(453.0,524.0){\rule[-0.200pt]{4.818pt}{0.400pt}}
\put(653.0,643.0){\rule[-0.200pt]{0.400pt}{8.913pt}}
\put(643.0,643.0){\rule[-0.200pt]{4.818pt}{0.400pt}}
\put(643.0,680.0){\rule[-0.200pt]{4.818pt}{0.400pt}}
\put(1014.0,711.0){\rule[-0.200pt]{0.400pt}{3.854pt}}
\put(1004.0,711.0){\rule[-0.200pt]{4.818pt}{0.400pt}}
\put(1004.0,727.0){\rule[-0.200pt]{4.818pt}{0.400pt}}
\put(1334.0,721.0){\rule[-0.200pt]{0.400pt}{4.095pt}}
\put(1324.0,721.0){\rule[-0.200pt]{4.818pt}{0.400pt}}
\put(1324.0,738.0){\rule[-0.200pt]{4.818pt}{0.400pt}}
\end{picture}
\caption{The ratio $R_A$ between isoscalar and isovector $g_A$ in 
VQCD and QCD are plotted against the dimensionless quark mass
$m_q a$ from the strange to the charm region. $\bigtriangleup $ indicates
the VQCD case, $\circ$/$\bullet$ indicates the C.I./ D.I. + C.I. in the 
QCD case. The dashed line is the $SU(6)$ prediction of 3/5.}
\end{figure}

  Now, we turn to valence QCD~\cite{ldd99}. This is the case where the
quark is forbidden to propagate backward in time so that there is no pair
creation and, as a result, only the valence $u$ and $d$ quarks are present in a
`nucleon'.  The motivation of studying this mutilated QCD is to establish
a connection between QCD and the valence quark model~\cite{ldd99}.
The same 100 gauge configurations used
for quenched QCD calculation are used for the valence QCD (VQCD) case. Since in
VQCD there is only C.I. (Fig. 1(a)), the $R_A$ ratio in Eq. (\ref{R_A}) becomes
\begin{equation}  \label{VR_A}
R_A =\frac{g_A^0}{g_A^3} (C. I.) 
= \frac{(\Delta u + \Delta d)(C.I.)}{(\Delta u - \Delta d)(C.I.)}. 
\end{equation}
The results are plotted in Fig. 2 as a function of the dimensionless 
quark mass $m_qa$
(with $\kappa$ = 0.162, 0.1615, 0.1610, 0.1590, and 0.1585) in comparison
with the QCD case. We see that, even for light quarks in 
the strange region ($m_q a \sim 0.07$), it is much closer to the valence 
prediction of 3/5, in contrast to the QCD calculation with C.I. alone.
This shows that VQCD indeed seems to confirm our expectation of the
valence quarks behavior, i.e. obeying the $SU(6)$ relation. The deviation
from the exact 3/5 prediction in Fig. 2 reflects the fact that there is still
a spin-spin interaction between the valence quarks as evidenced in the
$\vec{\sigma}\cdot\vec{B}$ term in the VQCD action with Pauli
spinors~\cite{ldd99}. Its effect, however, appears to be small.  This also
confirms our earlier assertion that the deviation of the C.I. of $R_A$ in QCD
($\circ$ in Fig. 2) is largely due to the the connected sea quark-antiquark
pairs.

With only the C.I., the $F_A/D_A$ ratio is related to VQCD $R_A$ in 
Eq. (\ref{VR_A})
\begin{equation}
\frac{F_A}{D_A} (C.I.) = \frac{1 + R_A}{3 - R_A},
\end{equation}    
From $R_A = 0.566(11)$ for the smallest quark mass ($\kappa = 0.162$), we
obtain $F_A/D_A = 0. 643(4)$. The fact that this is only slightly larger than
the QCD prediction of $0.60(2)$ for the \mbox{C.I.} (see Table 1) has to do
with the fact that the disconnected sea contribution is essentially independent
of flavor in our calculation, i.e. $\Delta u_s = \Delta d_s = \Delta s$
~\cite{dll95}.
As a result, $F_A$, $D_A$, and the $F_A/D_A$ ratio are identical with or 
without the sea quarks from the D.I. (see Table 1) and they do not
reflect the large disconnected sea effect due to the individual flavor.

\subsection{Scalar Matrix Elements, $R_S$, and $D_S/F_S$ }

  Similar situation exists for the scalar current matrix elements. It has
been suggested that the well-known $\pi N \sigma$ term 
($\sigma_{\pi N} =\hat{m}\langle N|\bar{u}u + \bar{d}d|N\rangle$ 
with $\hat{m} = (m_u + m_d)/2$) puzzle~\cite{che76,gls91} 
can be resolved because of the large OZI violating contribution from the
sea with a large $\bar{s}s$ content in the nucleon~\cite
{che76,gl82} such that $y= 2 \langle N|\bar{s}s|N\rangle/\langle N|
\bar{u}u + \bar{d}d |N\rangle \sim$ 0.2 -- 0.3. This has been verified
in lattice calculations~\cite{fko95a,dll96,guv98} which show that the D.I. is
$\sim$ 1.8 times of the C.I. (see Table 2)~\cite{dll96} and the $y$ ratio 
as large as $0.36 \pm 0.03$~\cite{dll96}.

 \begin{table}[ht]
\caption{Scalar contents, $\sigma_{\pi N}$, $F_S$, and $D_S$, in
comparison with phenomenology and quark model (QM). The
17.7 MeV in the last column is determined with the quark mass from the
 lattice calculation.}
\begin{tabular}{lllll}
 \multicolumn{1}{c}{} &\multicolumn{1}{c}{C. I.}&\multicolumn{1}{c}
 {C.I + D.I.}
 & \multicolumn{1}{c} {Phenomenology} & \multicolumn{1}{c} {QM}\\
 \hline
 $\langle p|\bar{u}u + \bar{d}d|p\rangle$ & 3.02(9) & 8.43(24) & & 
 $\leq$ 3 \\
 $\langle p|\bar{u}u - \bar{d}d|p\rangle$   & 0.63(9) &  & & $\leq$ 1 \\
 $\langle N|\bar{s}s|N\rangle$  & 1.53(7) &  & & 0 \\
 $F_S$ & 0.91(13) & 1.51(12) & 1.52 \cite{mmp87,oka96} --- 1.81\cite{gas81}
  & $\leq$ 1 \\
 $D_S$ & -0.28(10) & -0.88(28)& -0.52\cite{mmp87,oka96} --- -0.57\cite{gas81} 
  & 0  \\
  $\sigma_{\pi N}$ & 17.8(5) MeV & 49.7(2.6) MeV & 45 MeV \cite{gls91}
   & $\le 17.7$ MeV \\
  \hline
 \end{tabular}
 \end{table}

Unlike the case of the axial current matrix element, different flavors
contribute differently to the D.I. of the scalar matrix element --
$s$ contributes less than $u$ and $d$. As a result, 
the $SU(3)$ antisymmetric and symmetric parameters,
$F_S = (\langle p|\bar{u}u |p\rangle -\langle N|\bar{s}s|N\rangle)/2,
D_S = (\langle p|\bar{u}u |p\rangle - 2\langle p|\bar{d}d |p\rangle
 + \langle N|\bar{s}s|N\rangle)/2$
are strongly affected by the large D.I. part. We see from Table 2
that both $D_S$ and $F_S$ compare favorably with the
phenomenological values obtained from the SU(3) breaking pattern of
the octet baryon masses with either linear
\cite{mmp87,oka96} or quadratic mass relations \cite{gas81}.
This agreement is significantly improved from the valence
quark model which predicts $F_S < 1$ and $D_S = 0$ and also those of
the {\it C.I.} alone \cite{mmp87,oka96}.
The latter yields $F_S = 0.91(13)$ and $D_S = - 0.28(10)$ which are
only half of the phenomenological values \cite{mmp87,oka96,gas81}.
This again underscores the importance of the disconnected sea-quark
contributions.

Next, we address the effect of the connected sea quarks in the C.I.. Similar to
the ratio $R_A$ in the axial-vector case, we plot the ratio
\begin{eqnarray}    \label{R_S}
R_S &=&\frac{g_S^{I =0}}{g_S^{I=1}} = \frac{\langle p|\bar{u}u + 
\bar{d}d|p\rangle}{\langle p|\bar{u}u  - \bar{d}d|p\rangle} \nonumber \\
&=& \frac{(\langle p|\bar{u}u + \bar{d}d|p\rangle)(C. I.) +
(\langle p|\bar{u}u + \bar{d}d|p\rangle)(D. I.)}
{\langle p|\bar{u}u  - \bar{d}d|p\rangle}
\end{eqnarray}
as a function of the quark mass in Fig. 3.

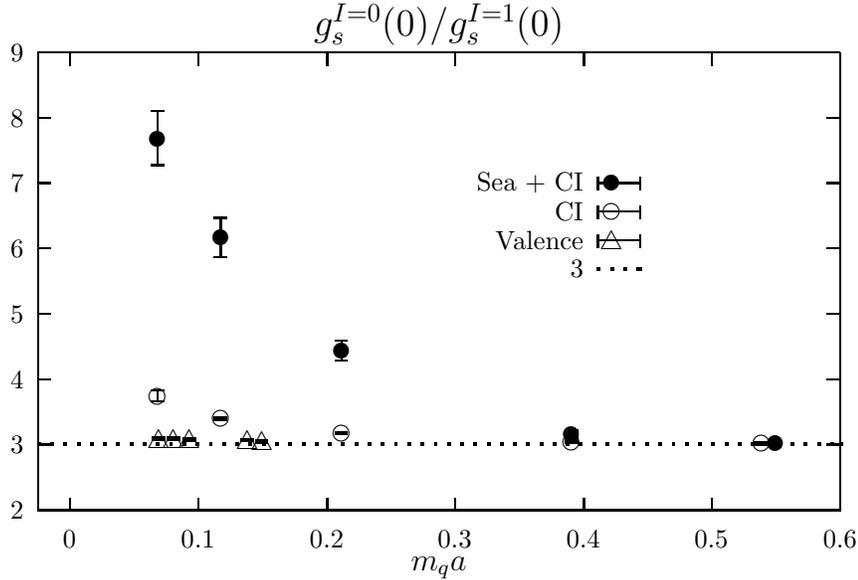
\begin{figure}[h]
\setlength{\unitlength}{0.240900pt}
\ifx\plotpoint\undefined\newsavebox{\plotpoint}\fi
\sbox{\plotpoint}{\rule[-0.200pt]{0.400pt}{0.400pt}}%
\begin{picture}(1500,900)(0,0)
\font\gnuplot=cmr10 at 10pt
\gnuplot
\sbox{\plotpoint}{\rule[-0.200pt]{0.400pt}{0.400pt}}%
\put(176.0,113.0){\rule[-0.200pt]{4.818pt}{0.400pt}}
\put(154,113){\makebox(0,0)[r]{2}}
\put(1416.0,113.0){\rule[-0.200pt]{4.818pt}{0.400pt}}
\put(176.0,216.0){\rule[-0.200pt]{4.818pt}{0.400pt}}
\put(154,216){\makebox(0,0)[r]{3}}
\put(1416.0,216.0){\rule[-0.200pt]{4.818pt}{0.400pt}}
\put(176.0,318.0){\rule[-0.200pt]{4.818pt}{0.400pt}}
\put(154,318){\makebox(0,0)[r]{4}}
\put(1416.0,318.0){\rule[-0.200pt]{4.818pt}{0.400pt}}
\put(176.0,421.0){\rule[-0.200pt]{4.818pt}{0.400pt}}
\put(154,421){\makebox(0,0)[r]{5}}
\put(1416.0,421.0){\rule[-0.200pt]{4.818pt}{0.400pt}}
\put(176.0,524.0){\rule[-0.200pt]{4.818pt}{0.400pt}}
\put(154,524){\makebox(0,0)[r]{6}}
\put(1416.0,524.0){\rule[-0.200pt]{4.818pt}{0.400pt}}
\put(176.0,627.0){\rule[-0.200pt]{4.818pt}{0.400pt}}
\put(154,627){\makebox(0,0)[r]{7}}
\put(1416.0,627.0){\rule[-0.200pt]{4.818pt}{0.400pt}}
\put(176.0,729.0){\rule[-0.200pt]{4.818pt}{0.400pt}}
\put(154,729){\makebox(0,0)[r]{8}}
\put(1416.0,729.0){\rule[-0.200pt]{4.818pt}{0.400pt}}
\put(176.0,832.0){\rule[-0.200pt]{4.818pt}{0.400pt}}
\put(154,832){\makebox(0,0)[r]{9}}
\put(1416.0,832.0){\rule[-0.200pt]{4.818pt}{0.400pt}}
\put(226.0,113.0){\rule[-0.200pt]{0.400pt}{4.818pt}}
\put(226,68){\makebox(0,0){0}}
\put(226.0,812.0){\rule[-0.200pt]{0.400pt}{4.818pt}}
\put(428.0,113.0){\rule[-0.200pt]{0.400pt}{4.818pt}}
\put(428,68){\makebox(0,0){0.1}}
\put(428.0,812.0){\rule[-0.200pt]{0.400pt}{4.818pt}}
\put(630.0,113.0){\rule[-0.200pt]{0.400pt}{4.818pt}}
\put(630,68){\makebox(0,0){0.2}}
\put(630.0,812.0){\rule[-0.200pt]{0.400pt}{4.818pt}}
\put(831.0,113.0){\rule[-0.200pt]{0.400pt}{4.818pt}}
\put(831,68){\makebox(0,0){0.3}}
\put(831.0,812.0){\rule[-0.200pt]{0.400pt}{4.818pt}}
\put(1033.0,113.0){\rule[-0.200pt]{0.400pt}{4.818pt}}
\put(1033,68){\makebox(0,0){0.4}}
\put(1033.0,812.0){\rule[-0.200pt]{0.400pt}{4.818pt}}
\put(1234.0,113.0){\rule[-0.200pt]{0.400pt}{4.818pt}}
\put(1234,68){\makebox(0,0){0.5}}
\put(1234.0,812.0){\rule[-0.200pt]{0.400pt}{4.818pt}}
\put(1436.0,113.0){\rule[-0.200pt]{0.400pt}{4.818pt}}
\put(1436,68){\makebox(0,0){0.6}}
\put(1436.0,812.0){\rule[-0.200pt]{0.400pt}{4.818pt}}
\put(176.0,113.0){\rule[-0.200pt]{303.534pt}{0.400pt}}
\put(1436.0,113.0){\rule[-0.200pt]{0.400pt}{173.207pt}}
\put(176.0,832.0){\rule[-0.200pt]{303.534pt}{0.400pt}}
\put(806,23){\makebox(0,0){{\large\bf $m_q a$}}}
\put(806,877){\makebox(0,0){{\Large\bf $g_s^{I=0}(0)/g_s^{I=1}(0)$}}}
\put(176.0,113.0){\rule[-0.200pt]{0.400pt}{173.207pt}}
\put(1033,627){\makebox(0,0)[r]{Sea + CI}}
\put(1077,627){\circle*{24}}
\put(364,697){\circle*{24}}
\put(463,541){\circle*{24}}
\put(653,364){\circle*{24}}
\put(1014,233){\circle*{24}}
\put(1334,218){\circle*{24}}
\put(1055.0,627.0){\rule[-0.200pt]{15.899pt}{0.400pt}}
\put(1055.0,617.0){\rule[-0.200pt]{0.400pt}{4.818pt}}
\put(1121.0,617.0){\rule[-0.200pt]{0.400pt}{4.818pt}}
\put(364.0,655.0){\rule[-0.200pt]{0.400pt}{20.476pt}}
\put(354.0,655.0){\rule[-0.200pt]{4.818pt}{0.400pt}}
\put(354.0,740.0){\rule[-0.200pt]{4.818pt}{0.400pt}}
\put(463.0,511.0){\rule[-0.200pt]{0.400pt}{14.695pt}}
\put(453.0,511.0){\rule[-0.200pt]{4.818pt}{0.400pt}}
\put(453.0,572.0){\rule[-0.200pt]{4.818pt}{0.400pt}}
\put(653.0,348.0){\rule[-0.200pt]{0.400pt}{7.468pt}}
\put(643.0,348.0){\rule[-0.200pt]{4.818pt}{0.400pt}}
\put(643.0,379.0){\rule[-0.200pt]{4.818pt}{0.400pt}}
\put(1014.0,227.0){\rule[-0.200pt]{0.400pt}{2.650pt}}
\put(1004.0,227.0){\rule[-0.200pt]{4.818pt}{0.400pt}}
\put(1004.0,238.0){\rule[-0.200pt]{4.818pt}{0.400pt}}
\put(1334.0,213.0){\rule[-0.200pt]{0.400pt}{2.409pt}}
\put(1324.0,213.0){\rule[-0.200pt]{4.818pt}{0.400pt}}
\put(1324.0,223.0){\rule[-0.200pt]{4.818pt}{0.400pt}}
\put(1033,582){\makebox(0,0)[r]{CI}}
\put(1077,582){\circle{24}}
\put(364,292){\circle{24}}
\put(463,257){\circle{24}}
\put(653,234){\circle{24}}
\put(1014,220){\circle{24}}
\put(1312,218){\circle{24}}
\put(1055.0,582.0){\rule[-0.200pt]{15.899pt}{0.400pt}}
\put(1055.0,572.0){\rule[-0.200pt]{0.400pt}{4.818pt}}
\put(1121.0,572.0){\rule[-0.200pt]{0.400pt}{4.818pt}}
\put(364.0,284.0){\rule[-0.200pt]{0.400pt}{4.095pt}}
\put(354.0,284.0){\rule[-0.200pt]{4.818pt}{0.400pt}}
\put(354.0,301.0){\rule[-0.200pt]{4.818pt}{0.400pt}}
\put(463.0,255.0){\rule[-0.200pt]{0.400pt}{0.964pt}}
\put(453.0,255.0){\rule[-0.200pt]{4.818pt}{0.400pt}}
\put(453.0,259.0){\rule[-0.200pt]{4.818pt}{0.400pt}}
\put(653.0,233.0){\rule[-0.200pt]{0.400pt}{0.482pt}}
\put(643.0,233.0){\rule[-0.200pt]{4.818pt}{0.400pt}}
\put(643.0,235.0){\rule[-0.200pt]{4.818pt}{0.400pt}}
\put(1014.0,218.0){\rule[-0.200pt]{0.400pt}{0.964pt}}
\put(1004.0,218.0){\rule[-0.200pt]{4.818pt}{0.400pt}}
\put(1004.0,222.0){\rule[-0.200pt]{4.818pt}{0.400pt}}
\put(1312.0,216.0){\rule[-0.200pt]{0.400pt}{0.964pt}}
\put(1302.0,216.0){\rule[-0.200pt]{4.818pt}{0.400pt}}
\put(1302.0,220.0){\rule[-0.200pt]{4.818pt}{0.400pt}}
\put(1033,537){\makebox(0,0)[r]{Valence}}
\put(1077,537){\raisebox{-.8pt}{\makebox(0,0){$\triangle$}}}
\put(366,225){\raisebox{-.8pt}{\makebox(0,0){$\triangle$}}}
\put(389,225){\raisebox{-.8pt}{\makebox(0,0){$\triangle$}}}
\put(414,224){\raisebox{-.8pt}{\makebox(0,0){$\triangle$}}}
\put(505,223){\raisebox{-.8pt}{\makebox(0,0){$\triangle$}}}
\put(528,221){\raisebox{-.8pt}{\makebox(0,0){$\triangle$}}}
\put(1055.0,537.0){\rule[-0.200pt]{15.899pt}{0.400pt}}
\put(1055.0,527.0){\rule[-0.200pt]{0.400pt}{4.818pt}}
\put(1121.0,527.0){\rule[-0.200pt]{0.400pt}{4.818pt}}
\put(366.0,223.0){\rule[-0.200pt]{0.400pt}{0.964pt}}
\put(356.0,223.0){\rule[-0.200pt]{4.818pt}{0.400pt}}
\put(356.0,227.0){\rule[-0.200pt]{4.818pt}{0.400pt}}
\put(389.0,223.0){\rule[-0.200pt]{0.400pt}{0.964pt}}
\put(379.0,223.0){\rule[-0.200pt]{4.818pt}{0.400pt}}
\put(379.0,227.0){\rule[-0.200pt]{4.818pt}{0.400pt}}
\put(414.0,222.0){\rule[-0.200pt]{0.400pt}{0.964pt}}
\put(404.0,222.0){\rule[-0.200pt]{4.818pt}{0.400pt}}
\put(404.0,226.0){\rule[-0.200pt]{4.818pt}{0.400pt}}
\put(505.0,221.0){\rule[-0.200pt]{0.400pt}{0.964pt}}
\put(495.0,221.0){\rule[-0.200pt]{4.818pt}{0.400pt}}
\put(495.0,225.0){\rule[-0.200pt]{4.818pt}{0.400pt}}
\put(528.0,219.0){\rule[-0.200pt]{0.400pt}{0.964pt}}
\put(518.0,219.0){\rule[-0.200pt]{4.818pt}{0.400pt}}
\put(518.0,223.0){\rule[-0.200pt]{4.818pt}{0.400pt}}
\sbox{\plotpoint}{\rule[-0.500pt]{1.000pt}{1.000pt}}%
\put(1033,492){\makebox(0,0)[r]{3}}
\multiput(1055,492)(20.756,0.000){4}{\usebox{\plotpoint}}
\put(1121,492){\usebox{\plotpoint}}
\put(176,216){\usebox{\plotpoint}}
\multiput(176,216)(20.756,0.000){61}{\usebox{\plotpoint}}
\put(1436,216){\usebox{\plotpoint}}
\end{picture}
\caption{The ratio $R_S$ between isoscalar and isovector scalar charge  
in QCD (Eq. (\ref{R_S})) and VQCD (Eq. (\ref{VR_S}) are plotted against the 
dimensionless quark mass $m_q a$ from the strange to the charm region. 
$\circ$/$\bullet$ indicates the C.I./ D.I. + C.I. in the
QCD case and $\triangle$ indicates
the VQCD case. The dashed line is the valence quark model
 prediction of 3.}
\end{figure}

The dotted line is the valence quark model prediction of 3 for both
the non-relativistic and relativistic cases. Again for heavy quarks
(i.e. $\kappa \geq 0.133$ or $ma \geq 0.4$ in Fig. 3), we see that 
the ratio $R_S$ is 3 irrespective whether the D.I. is included	
(shown as $\bullet$ in Fig. 3) or not (C.I. alone is indicated as
$\circ$). As for light quarks, the full result
(C.I. + D.I.) is much larger than 3 largely due to the large
contribution in the D.I. (NB. Table 2 lists the results at
the chiral limit).  Even for the C.I. alone, 
$R_S$ still overshoots 3. As we shall see, this is again caused
by the connected sea quarks. For VQCD, the $R_S$ ratio becomes

\begin{equation}  \label{VR_S}
R_S = \frac{(\langle p|\bar{u}u + \bar{d}d|p\rangle)(C.I.)}
{(\langle p|\bar{u}u  - \bar{d}d|p\rangle)(C.I.)}.
\end{equation}
We see in Fig. 3
that the ratios (denoted by $\triangle$) for the light quarks are
approaching  the valence quark
prediction of 3. This again confirms that the deviation of the C.I. result
in QCD is primarily due to the Z-graphs with connected sea quarks and
antiquarks. When they are eliminated in VQCD, $R_S$ becomes close to the
$SU(6)$ relation.

The $D_S/F_S$ ratio in VQCD is
\begin{equation}
D_S/F_S (C. I.) = \frac{3 - R_S}{1 + R_S},
\end{equation} 
From $R_S = 3.086(19)$ for the smallest quark mass ($\kappa = 0.162$), we
obtain $D_S/F_S = - 0.021(4)$ which is close to zero as in the valence quark 
picture  (Table 2) and differs from the lattice QCD calculation of - 0.58(18) 
(D.I. + C.I.) and -0.31(11) (C.I.) (see Table 2) by a large margin. 


\subsection{Strangeness Magnetic Moment and Neutron to Proton Magnetic Moment
Ratio}

After having established the importance of the strangeness sea effects in
the axial and scalar matrix elements, one
would naturally ask what happens to the vector current matrix elements,
especially the neutron to proton magnetic
moment ratio $\mu_n/\mu_p$. How much will the sea affect the 
ratio and in what way? After all, the $\mu_n/\mu_p$ ratio was well
predicted by the valence picture -- a celebrated
defining success of the $SU(6)$ symmetry. 

The strangeness contribution to the proton electric and magnetic form factors
$G_E^s(q^2)$ and $G_M^s(q^2)$ can be extracted from parity violating
asymmetry measured in the polarized electron proton scattering~\cite{bec00}.
The first result coming out of the SAMPLE experiment~\cite{sample97} gives the
strangeness magnetic form factor $G_M^s (Q^2 = 0.1 {\rm GeV}^2) = 0.23 \pm 0.37
\pm 0.15 \pm 0.19$ n.m.  Combined with data from the deuteron, the new result is
closer to zero with a relatively larger error~\cite{bec00}.  Our lattice
analysis gives $G_M^s (Q^2 = 0) = - 0.36 \pm 0.20$ n.m.~\cite{dlw98}. With an
unbiased subtraction to reduce the noise, the new analysis gives $G_M^s (Q^2 =
0) = - 0.28\pm 0.10$ n.m.~\cite{mat00}. Although it is consistent with the
SAMPLE experiment within error bars, the central value may differ by a sign.
Theoretically, very little is known about the strangeness contribution to the
EM form factor of the nucleon.  Even the sign of the strangeness contribution
to the magnetic moments is uncertain.  Different model gives different
prediction, including its sign~\cite{lei96}.  Hopefully, the experimental
results will be clearer when the approved experiments at CEBAF (Hall C
experiment 91-017)~\cite{ceb91} start to produce results.  We have also
calculated the strange electric form factor $G_E^s (q^2)$ and found it to be
small and positive.  This is going to be a prediction which can be checked by
the CEBAF experiments~\cite{ceb91}.

According to the lattice calculation~\cite{dlw98}, the $u$ and $d$ contributions
are $\sim$ 80\% larger, $G_M^{u,d}(0) (D. I.) = -0.65 \pm 0.30$. However, their
net contribution to the proton and neutron magnetic moment

\begin{eqnarray}  \label{mudi}
\mu (D. I.) &= &(2/3 G_M^u(0) (D. I.) -1/3 G_M^d(0) (D. I.) -1/3
G_M^s(0)) \mu_N  \nonumber \\
&= & -0.097 \pm 0.037 \mu_N 
\end{eqnarray}
becomes smaller due the cancellation of the quark charges of $u, d,$ and $s$. 
The result still holds within error after the unbiased subtraction~\cite{mat00}.

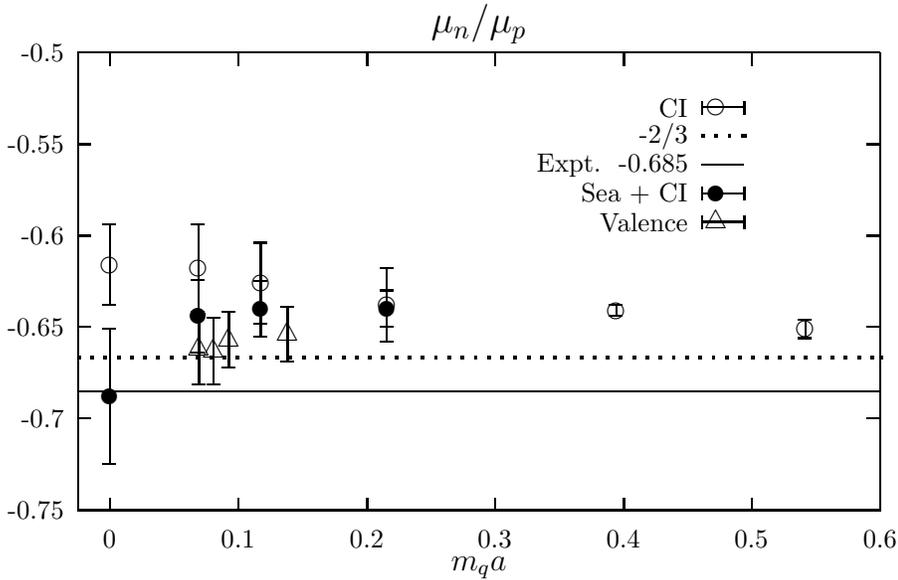
\begin{figure}[h]
\setlength{\unitlength}{0.240900pt}
\ifx\plotpoint\undefined\newsavebox{\plotpoint}\fi
\sbox{\plotpoint}{\rule[-0.200pt]{0.400pt}{0.400pt}}%
\begin{picture}(1500,900)(0,0)
\font\gnuplot=cmr10 at 10pt
\gnuplot
\sbox{\plotpoint}{\rule[-0.200pt]{0.400pt}{0.400pt}}%
\put(176.0,113.0){\rule[-0.200pt]{4.818pt}{0.400pt}}
\put(154,113){\makebox(0,0)[r]{-0.75}}
\put(1416.0,113.0){\rule[-0.200pt]{4.818pt}{0.400pt}}
\put(176.0,257.0){\rule[-0.200pt]{4.818pt}{0.400pt}}
\put(154,257){\makebox(0,0)[r]{-0.7}}
\put(1416.0,257.0){\rule[-0.200pt]{4.818pt}{0.400pt}}
\put(176.0,401.0){\rule[-0.200pt]{4.818pt}{0.400pt}}
\put(154,401){\makebox(0,0)[r]{-0.65}}
\put(1416.0,401.0){\rule[-0.200pt]{4.818pt}{0.400pt}}
\put(176.0,544.0){\rule[-0.200pt]{4.818pt}{0.400pt}}
\put(154,544){\makebox(0,0)[r]{-0.6}}
\put(1416.0,544.0){\rule[-0.200pt]{4.818pt}{0.400pt}}
\put(176.0,688.0){\rule[-0.200pt]{4.818pt}{0.400pt}}
\put(154,688){\makebox(0,0)[r]{-0.55}}
\put(1416.0,688.0){\rule[-0.200pt]{4.818pt}{0.400pt}}
\put(176.0,832.0){\rule[-0.200pt]{4.818pt}{0.400pt}}
\put(154,832){\makebox(0,0)[r]{-0.5}}
\put(1416.0,832.0){\rule[-0.200pt]{4.818pt}{0.400pt}}
\put(226.0,113.0){\rule[-0.200pt]{0.400pt}{4.818pt}}
\put(226,68){\makebox(0,0){0}}
\put(226.0,812.0){\rule[-0.200pt]{0.400pt}{4.818pt}}
\put(428.0,113.0){\rule[-0.200pt]{0.400pt}{4.818pt}}
\put(428,68){\makebox(0,0){0.1}}
\put(428.0,812.0){\rule[-0.200pt]{0.400pt}{4.818pt}}
\put(630.0,113.0){\rule[-0.200pt]{0.400pt}{4.818pt}}
\put(630,68){\makebox(0,0){0.2}}
\put(630.0,812.0){\rule[-0.200pt]{0.400pt}{4.818pt}}
\put(831.0,113.0){\rule[-0.200pt]{0.400pt}{4.818pt}}
\put(831,68){\makebox(0,0){0.3}}
\put(831.0,812.0){\rule[-0.200pt]{0.400pt}{4.818pt}}
\put(1033.0,113.0){\rule[-0.200pt]{0.400pt}{4.818pt}}
\put(1033,68){\makebox(0,0){0.4}}
\put(1033.0,812.0){\rule[-0.200pt]{0.400pt}{4.818pt}}
\put(1234.0,113.0){\rule[-0.200pt]{0.400pt}{4.818pt}}
\put(1234,68){\makebox(0,0){0.5}}
\put(1234.0,812.0){\rule[-0.200pt]{0.400pt}{4.818pt}}
\put(1436.0,113.0){\rule[-0.200pt]{0.400pt}{4.818pt}}
\put(1436,68){\makebox(0,0){0.6}}
\put(1436.0,812.0){\rule[-0.200pt]{0.400pt}{4.818pt}}
\put(176.0,113.0){\rule[-0.200pt]{303.534pt}{0.400pt}}
\put(1436.0,113.0){\rule[-0.200pt]{0.400pt}{173.207pt}}
\put(176.0,832.0){\rule[-0.200pt]{303.534pt}{0.400pt}}
\put(806,23){\makebox(0,0){{\large\bf $m_q a$}}}
\put(806,877){\makebox(0,0){{\Large\bf $\mu_n/\mu_p$}}}
\put(176.0,113.0){\rule[-0.200pt]{0.400pt}{173.207pt}}
\put(1134,746){\makebox(0,0)[r]{CI}}
\put(1178,746){\circle{24}}
\put(226,498){\circle{24}}
\put(365,493){\circle{24}}
\put(463,470){\circle{24}}
\put(661,435){\circle{24}}
\put(1021,426){\circle{24}}
\put(1318,398){\circle{24}}
\put(1156.0,746.0){\rule[-0.200pt]{15.899pt}{0.400pt}}
\put(1156.0,736.0){\rule[-0.200pt]{0.400pt}{4.818pt}}
\put(1222.0,736.0){\rule[-0.200pt]{0.400pt}{4.818pt}}
\put(226.0,435.0){\rule[-0.200pt]{0.400pt}{30.594pt}}
\put(216.0,435.0){\rule[-0.200pt]{4.818pt}{0.400pt}}
\put(216.0,562.0){\rule[-0.200pt]{4.818pt}{0.400pt}}
\put(365.0,424.0){\rule[-0.200pt]{0.400pt}{33.244pt}}
\put(355.0,424.0){\rule[-0.200pt]{4.818pt}{0.400pt}}
\put(355.0,562.0){\rule[-0.200pt]{4.818pt}{0.400pt}}
\put(463.0,406.0){\rule[-0.200pt]{0.400pt}{30.594pt}}
\put(453.0,406.0){\rule[-0.200pt]{4.818pt}{0.400pt}}
\put(453.0,533.0){\rule[-0.200pt]{4.818pt}{0.400pt}}
\put(661.0,378.0){\rule[-0.200pt]{0.400pt}{27.703pt}}
\put(651.0,378.0){\rule[-0.200pt]{4.818pt}{0.400pt}}
\put(651.0,493.0){\rule[-0.200pt]{4.818pt}{0.400pt}}
\put(1021.0,418.0){\rule[-0.200pt]{0.400pt}{4.095pt}}
\put(1011.0,418.0){\rule[-0.200pt]{4.818pt}{0.400pt}}
\put(1011.0,435.0){\rule[-0.200pt]{4.818pt}{0.400pt}}
\put(1318.0,383.0){\rule[-0.200pt]{0.400pt}{6.986pt}}
\put(1308.0,383.0){\rule[-0.200pt]{4.818pt}{0.400pt}}
\put(1308.0,412.0){\rule[-0.200pt]{4.818pt}{0.400pt}}
\sbox{\plotpoint}{\rule[-0.500pt]{1.000pt}{1.000pt}}%
\put(1134,701){\makebox(0,0)[r]{-2/3}}
\multiput(1156,701)(20.756,0.000){4}{\usebox{\plotpoint}}
\put(1222,701){\usebox{\plotpoint}}
\put(176,353){\usebox{\plotpoint}}
\multiput(176,353)(20.756,0.000){61}{\usebox{\plotpoint}}
\put(1436,353){\usebox{\plotpoint}}
\sbox{\plotpoint}{\rule[-0.200pt]{0.400pt}{0.400pt}}%
\put(1134,656){\makebox(0,0)[r]{Expt.~~-0.685}}
\put(1156.0,656.0){\rule[-0.200pt]{15.899pt}{0.400pt}}
\put(176,300){\usebox{\plotpoint}}
\put(176.0,300.0){\rule[-0.200pt]{303.534pt}{0.400pt}}
\put(1134,611){\makebox(0,0)[r]{Sea + CI}}
\put(1178,611){\circle*{24}}
\put(226,291){\circle*{24}}
\put(365,418){\circle*{24}}
\put(463,429){\circle*{24}}
\put(661,429){\circle*{24}}
\put(1156.0,611.0){\rule[-0.200pt]{15.899pt}{0.400pt}}
\put(1156.0,601.0){\rule[-0.200pt]{0.400pt}{4.818pt}}
\put(1222.0,601.0){\rule[-0.200pt]{0.400pt}{4.818pt}}
\put(226.0,185.0){\rule[-0.200pt]{0.400pt}{51.312pt}}
\put(216.0,185.0){\rule[-0.200pt]{4.818pt}{0.400pt}}
\put(216.0,398.0){\rule[-0.200pt]{4.818pt}{0.400pt}}
\put(365.0,360.0){\rule[-0.200pt]{0.400pt}{27.703pt}}
\put(355.0,360.0){\rule[-0.200pt]{4.818pt}{0.400pt}}
\put(355.0,475.0){\rule[-0.200pt]{4.818pt}{0.400pt}}
\put(463.0,386.0){\rule[-0.200pt]{0.400pt}{20.958pt}}
\put(453.0,386.0){\rule[-0.200pt]{4.818pt}{0.400pt}}
\put(453.0,473.0){\rule[-0.200pt]{4.818pt}{0.400pt}}
\put(661.0,401.0){\rule[-0.200pt]{0.400pt}{13.731pt}}
\put(651.0,401.0){\rule[-0.200pt]{4.818pt}{0.400pt}}
\put(651.0,458.0){\rule[-0.200pt]{4.818pt}{0.400pt}}
\put(1134,566){\makebox(0,0)[r]{Valence}}
\put(1178,566){\makebox(0,0){$\triangle$}}
\put(366,366){\makebox(0,0){$\triangle$}}
\put(389,363){\makebox(0,0){$\triangle$}}
\put(413,380){\makebox(0,0){$\triangle$}}
\put(505,389){\makebox(0,0){$\triangle$}}
\put(1156.0,566.0){\rule[-0.200pt]{15.899pt}{0.400pt}}
\put(1156.0,556.0){\rule[-0.200pt]{0.400pt}{4.818pt}}
\put(1222.0,556.0){\rule[-0.200pt]{0.400pt}{4.818pt}}
\put(366.0,311.0){\rule[-0.200pt]{0.400pt}{26.499pt}}
\put(356.0,311.0){\rule[-0.200pt]{4.818pt}{0.400pt}}
\put(356.0,421.0){\rule[-0.200pt]{4.818pt}{0.400pt}}
\put(389.0,311.0){\rule[-0.200pt]{0.400pt}{25.054pt}}
\put(379.0,311.0){\rule[-0.200pt]{4.818pt}{0.400pt}}
\put(379.0,415.0){\rule[-0.200pt]{4.818pt}{0.400pt}}
\put(413.0,337.0){\rule[-0.200pt]{0.400pt}{20.958pt}}
\put(403.0,337.0){\rule[-0.200pt]{4.818pt}{0.400pt}}
\put(403.0,424.0){\rule[-0.200pt]{4.818pt}{0.400pt}}
\put(505.0,346.0){\rule[-0.200pt]{0.400pt}{20.717pt}}
\put(495.0,346.0){\rule[-0.200pt]{4.818pt}{0.400pt}}
\put(495.0,432.0){\rule[-0.200pt]{4.818pt}{0.400pt}}
\end{picture}
\caption{The ratio of neutron to proton magnetic moment  
$\mu_n/\mu_p$ is plotted against the dimensionless
quark mass. $\circ$ indicates the C.I. result only and  $\bullet$ 
shows the full result with both C.I. and D.I. and $\bigtriangleup $ 
indicates the ratio in the VQCD case.
The solid line is the valence quark model prediction of - 2/3 and the
dashed line is the experimental result of - 0.685.} 
\end{figure}

As illustrated in Fig. 4, where the neutron to proton magnetic moment ratio 
is plotted against the quark mass, this small $SU(6)$-breaking sea quark 
effect is further nullified by the connected sea effect~\cite{dlw98}. As a
result,  the $\mu_n/\mu_p$ ratio for the combined C.I. and D.I. comes to 
$-0.68 \pm 0.04$ at the chiral limit. This is quite consistent with the 
experimental value of $-0.685$. Barring any as yet unknown symmetry principle,
this cancellation between the connected and disconnected sea contributions
is probably accidental and in stark contrast
to the $\pi N \sigma$ term and flavor-singlet $g_A^0$ where the
connected and disconnected sea effects add up to enhance the $SU(6)$ breaking.

Also shown in Fig. 4 are results of VQCD  (indicated as 
$\bigtriangleup$) which are very close to the $SU(6)$
value of - 2/3 (the result for the smallest quark mass case is - 0.662(22)), 
indirectly verifying the connected sea effect of QCD ($\circ$ for the C.I. 
in Fig. 4) which shows a 2.5 $\sigma$ departure from - 2/3 at the chiral limit.
If there is any deviation of the VQCD from - 2/3, it should be due to the 
residual spin-spin interaction between the quarks in the baryon. Given the 
size of the error in our present results, we cannot make a definite 
conclusion on this aspect.

\section{Hyperon-Antihyperon Production in Nucleus-Nucleus Collisions}

We see from the previous section that there are large strangeness contents
in the nucleon. The strangeness quark spin ($\Delta s = -0.12(1)$) is 12\% of
the total nucleon angular momentum, albeit in the opposite direction.
The scalar matrix element for the strange is also large. With the ratio
$\langle N|\bar{s}s|N\rangle/\langle N| \bar{u}u + \bar{d}d |N\rangle = 
0.18 \pm 0.02$ from the lattice calculation, the strangeness accounts for
$\sim $ 18\% of the nucleon content in the scalar channel. This strongly
suggests that one needs to take the $s\bar{s}$ content in the form of
$\bar{s} s, \bar{s}\gamma_{\mu}\gamma_5 s, \bar{s}\gamma_{\mu} s, 
\bar{s}\gamma_5 s$, etc. in the nucleon wavefunction in additional to
the valence $u$ and $d$ quark wavefunction.

We heard in this conference~\cite{talks2000} that various experiments
at CERN SPS revealed an enhancement of strange particle production from
proton-proton and proton-nucleus to nucleus-nucleus collision~\cite{WA97,NA49}.
Upon analyzing their Pb-Pb data and comparing to p-Be and p-Pb data, the
WA97 collaboration found an enhancement in hyper-antihyperon ($\Lambda, 
\bar{\Lambda}, \Xi^-, \bar{\Xi}^+, \Omega^-$, and $\bar{\Omega}^+$) yields
per participant nucleon (wounded nucleon participating in the collions as
estimated from the Glauber model)~\cite{WA97}. Furthermore, the enhancement
is shown to increase with the valence strangeness in the produced hyperon
(antihyperon)~\cite{WA97}. Besides the scenario of equilibrated quark
gluon plasma phase~\cite{rm82,bhs99}, several mechanisms have been
considered theoretically in view of this enahancement. They include the
baryon junction exchange mechanism~\cite{vg99}, the diquark breaking
mechanism~\cite{cs99}, and the strong color field effect~\cite{bgs00}.

In view of the existence of non-negligible strangeness content in the nucleon,
it seems to me that one should take it into account in the hyperon-antihyperon
production from the proton-nucleus and nucleus-nucleus collisions at
high energies. Since the sea-quark content is relatively small compared to the
valence, it should be a good approximation to consider one pair of
$u\bar{u}, d\bar{d}$ and $s\bar{s}$ in the higher Fock space. In this case,
we can point out a difference between the p-nucleus and nucleus-nucleus
collisions. In the p-nucleus collsion, the $\Omega^- - \bar{\Omega}^+$ 
production can only come from the pair-creation via perturbative gluons which
is characterized by the coupling constant $\alpha_s$ at the SPS energy or
string breaking. On the other hand, the nucleus-nucleus collision can
involve additional mechanisms. For example, the pre-existing $s\bar{s}$
pairs in three nucleons can coalesce to form an $\Omega$ or $\bar{\Omega}$
during the collision. A combination of partial coalescence and
pair-creation is also possible. This is characterized by the small $s\bar{s}$
component in the nucleon wavefuntion, but it can be a competing mechanism if
the amplitue is comparable to that of the creation of 3 $s\bar{s}$ pairs via
gluons.
Furthermore, due to the combinatorics, one expects that the yield due to the
dissociation-recombination mechanism from the pre-existing sea pairs will grow
with the number of participant nucleons relative to that of the
pair-creation mechanism. 

To conclude, we have learned that there are surprisingly large 
strangeness contents in the nucleon in various channels, such as in the
the axail ($\bar{s}\gamma_{\mu}\gamma_5 s$) and scalar ($\bar{s}s$) matrix
elements. It would be interesting to find out what role they play in the hyper-
antihyperon productions in heavy-ion collisions.

\end{document}